\documentclass[usenatbib]{mn2e}
\usepackage{epsfig} 
\usepackage{graphicx}

\usepackage{url} 
\usepackage{amsmath}
\usepackage{graphicx}
\usepackage{ulem}

\usepackage[usenames]{color}
\definecolor{purple}{RGB}{160,32,240}
\definecolor{dgreen}{RGB}{0,128,0}

\def\dout{\bgroup
 \markoverwith{\lower-0.2ex\hbox
 {\kern-.03em\vbox{\hrule width.2em\kern0.45ex\hrule}\kern-.03em}}%
 \ULon}
\MakeRobust\dout

\newcommand{\cs}{c_{\rm s}}

\newcommand{\sigSB}{\sigma_{\rm SB}}
\newcommand{\arad}{a_{\rm rad}}

\newcommand{\K}{\mathrm{K}}

\newcommand{\dd}{\partial}

\newcommand{\s}{\mathrm{s}}
\newcommand{\erg}{\mathrm{erg}}

\newcommand{\yr}{\mathrm{yr}}

\newcommand{\Gyr}{\mathrm{Gyr}}

\newcommand{\Hz}{\mathrm{Hz}}

\newcommand{\pc}{\mathrm{pc}}

\newcommand{\Msol}{\mathrm{M}_{\odot}}

\newcommand{\apj}{ApJ}
\newcommand{\apjl}{ApJ}

\newcommand{\aap}{A$\&$A}
\newcommand{\CQG}{Class. Quantum Gravity}
\newcommand{\araa}{ARAA}

\newcommand{\mnras}{MNRAS}
\newcommand{\pasj}{PASJ}
\newcommand{\prd}{PRD}
\newcommand{\aj}{AJ}
\newcommand{\nat}{Nature}
\newcommand{\nar}{New Astron. Rev.}

\newcommand{\ssr}{{Space~Sci.~Rev.}}%
\newcommand{\memsai}{{Mem.~Soc.~Astron.~Italiana}}
\def\gta{\mathrel{\spose{\lower 3pt\hbox{$\mathchar"218$}}
        \raise 2.0pt\hbox{$\mathchar"13E$}}}

    \def\dd{\partial}

    \def\beq{\begin{equation} }
    \def\eeq{\end{equation} }
    \def\spose#1{\hbox to 0pt{#1\hss}}
    \def\ltsim{\mathrel{\spose{\lower.5ex\hbox{$\mathchar"218$}}
     \raise.4ex\hbox{$\mathchar"13C$}}}

\title[Recurring flares from supermassive black hole binaries]
{Recurring flares from supermassive black hole binaries:
implications for tidal disruption candidates and OJ 287
}
\author[T.L. Tanaka]
{
Takamitsu L. Tanaka$^{1}$\thanks{E-mail: taka@mpa-garching.mpg.de}
\\$^{1}$Max Planck Institute for Astrophysics, Karl-Schwarzschild-Strasse 1, 85741 Garching, 
Bavaria, Germany
}

\begin{document}

\maketitle

\label{firstpage}
\begin{abstract}
I discuss the possibility that accreting supermassive black hole (SMBH) binaries
with sub-parsec separations produce periodically recurring luminous outbursts
that interrupt periods of relative quiescence.
This hypothesis is motivated by two characteristics found
generically in simulations of binaries embedded in prograde accretion discs:
(i) the formation of a central, low-density cavity around the binary,
and (ii) the leakage of gas into this cavity,
occurring once per orbit via discrete streams on nearly radial trajectories.
The first feature would reduce the emergent optical/UV flux
of the system relative to active galactic nuclei
powered by single SMBHs, while the second can trigger
quasiperiodic fluctuations in luminosity.
I argue that the quasiperiodic accretion signature
may be much more dramatic than previously thought,
because the infalling gas streams can strongly shock-heat
via self-collision and tidal compression,
thereby enhancing viscous accretion.
Any optically thick gas that is circularized about either SMBH
can accrete before the next pair of streams is deposited,
fueling transient, luminous flares that recur every orbit.
Due to the diminished flux in between accretion episodes,
such cavity-accretion flares could plausibly be mistaken
for the tidal disruptions of stars in quiescent nuclei.
The flares could be distinguished from tidal disruption events
if their quasiperiodic recurrence is observed,
or if they are produced by very massive ($\gta 10^9~\Msol$) SMBHs
that cannot disrupt solar-type stars.
They may be discovered serendipitously 
in surveys such as \textit{LSST} or \textit{eROSITA}.
I present a heuristic toy model as a proof of concept
for the production of cavity-accretion flares,
and generate mock light curves and specta.
I also apply the model to the active galaxy OJ 287,
whose production of quasiperiodic pairs of optical flares
has long fueled speculation that it hosts a SMBH binary.
\end{abstract}

\section{Introduction}
\label{sec:intro}

Given that massive galaxies assemble through hierarchical mergers
and that virtually all of them host a supermassive black hole (SMBH) in their nuclei
\citep[see, e.g., the review by][and references therein]{FerrareseFord05},
it is inevitable that they spend a fraction of their time hosting two or more SMBHs.
Such a configuration will eventually result in the formation of a bound SMBH binary,
whose orbit becomes increasingly compact as it loses its orbital angular momentum
and energy---(I) first through dynamical friction with the stellar and gaseous background,
(II) then through three-body interactions with stars and/or
tidal interactions with a gaseous accretion disc,
and (III) finally, at separations well below a parsec,
being driven to merger through the emission of gravitational waves \citep{Begelman+80,Yu02,HKM09,ColpiDotti11}.

If identified, such systems could unveil
invaluable clues about the physics of accretion processes
in active galactic nuclei (AGN)
and the co-evolution of galaxies and their nuclear SMBHs.
An especially tantalizing possibility is that of observing near-merger
binaries concomitantly through both electromagnetic
and gravitational-wave signatures.
Not only would such multi-messenger studies
provide unprecedented independent determination of the
masses and spins of the central AGN engines,
they can also be used to probe the cosmic expansion history
(\citealt{HH05}, \citealt{Kocsis+06}, \citealt{Bloom+09}; see \citealt{Schnittman11}
and \citealt{TanakaHaiman13} for an overview of proposed electromagnetic counterparts).

SMBH binaries, however, have thus far mostly eluded discovery.
Although there are now numerous examples of dual and even triple AGN
at kpc-scale separations
\citep[e.g.:][]{Komossa+03,Bianchi+08,Comerford+09b,Green+10,Liu+10,Barth+08,Liu+11_1},
there is only one unambiguous example of a
gravitationally bound binary,
which has a projected separation of   7.3 pc \citep[the radio galaxy 0402+379;][]{Rodriguez+06}.
In searches for periodic variability in AGN,
a binary explanation has been ruled out for dozens of candidates
\citep{HalpFilipp88, Eracleous+97,HalpErac00, EracHalp03,Chornock+09}.
While there are several examples of AGN with double-peaked broad lines \citep{Gaskell83, Peterson+87, Gaskell96, BorLau09},
shifted lines \citep{Bogdan+09b, Dotti+09},
or suggestive radio lobe morphology \citep{Roos+93,GopalKrishna+03,Liu+03},
such features have alternate explanations that do not require the presence of a binary
\citep[see the review by][]{Komossa06}.
The blazar OJ 287 (a.k.a. EGO 0851+202), which we will revisit in \S\ref{subsec:OJ287},
has been suggested to host a binary SMBH because it
produces pairs of optical flares at regular intervals
of approximately 11-12 years \citep{Sillanpaa+88,Valtonen+08}.

The dearth of observational evidence is not surprising, for several reasons.
First, the detection of SMBH binaries must lean heavily
on indirect observations of kinematic signatures over long periods of time,
since resolving images of sub-parsec binaries 
is intractable in all wavelengths
but radio, where it is feasible only for relatively nearby sources
(e.g., for the case of 0402+379, which has a redshift of $z\approx 0.055$,
a projected distance of $1~{\rm pc}$ corresponds to $\sim 1~{\rm mas}$ on the sky). Second, because gravitational wave emission
and accretion disc torquing accelerate the orbital evolution of compact binaries,
the most compact binaries---whose kinematic
signatures would be the easiest to detect---should also be the rarest \citep[e.g.][and references therein]{HKM09}.
Third, the emission features of an accreting SMBH binary
may deviate significantly from those of AGN hosting a single SMBH,
potentially causing such systems to be missed in AGN surveys.
For example, many theoretical studies suggest that if the binary's SMBH
masses are similar and its orbit is prograde with its accretion disc,
then the binary's tidal torques can suppress the accretion rate in its vicinity
and form a low-density, central, circumbinary cavity in the disc
\citep[e.g.:][]{ArtLub94, Hayasaki+07, MM08, Farris+12}.
Such a cavity would result in reduced emission in soft X-ray,
UV and even optical frequencies compared to typical AGN \citep{MP05, TMH12, GultekinMiller12}.

Numerical simulations of  such circumbinary discs agree that
the suppression of the  accretion rate into the cavity is not absolute.
Once per orbital period, gas leaks into the cavity in discrete,
elongated streams \citep[e.g.:][]{ArtLub96, Cuadra+09, D'Orazio+12},
a portion of which will accrete onto one or both SMBHs.
The observable signature of the quasiperiodic streams depends on
the timescales on which they are accreted by the SMBHs.
If the accretion timescale of an individual stream
is long, then the leaked gas will form a small accretion disc
around one or both SMBHs (e.g., Hayasaki et al. 2008). This quasiperiodic accretion feature has long been suggested to trigger
corresponding variability in AGN emission,
and invoked to explain objects like OJ 287 as SMBH binaries \citep{ArtLub96, Hayasaki+12}. 

In this work, I consider the possibility that the streams are instead
accreted on timescales shorter than the binary's orbital period.
This is motivated by the fact that because the streams are elongated
and enter on nearly radial orbits, they are prone to shock-heating
by tidal elongation and self-crossing, as well as by tidal compression.
Because the streams are marginally bound initially, the fraction of orbital
kinetic energy that can be converted to thermal energy is large.
Any post-shock gas that is bound to either SMBH will be much hotter
than the gas in standard AGN solutions at similar radii,
and thus have much shorter viscous diffusion timescales if it is optically thick.
I show that for a wide range of plausible parameters,
the shock-heated accretion flow will be consumed by either SMBH
on timescales shorter than the binary orbital period---i.e., before the next pair
of streams enters the cavity.
In this scenario, the quasiperiodic leakage of gas into the cavity
fuels recurrent, transient flares, with UV-bright
spectral energy distributions (SEDs) similar to luminous AGN.
The flares would recur once or twice per binary orbit,
depending on whether one or both SMBHs accrete.

Because the central cavity causes the disc to be
dim in between the flares, an individual flare
may resemble the tidal disruption of a star by a SMBH
(tidal disruption event, hereafter TDE; \citealt{Rees88}).
Cavity-accretion flares from SMBH binaries may be observed,
potentially in large numbers,
by deep surveys that revisit the same portions of the sky
multiple times,
such as  \textit{LSST}\footnote{
Large Synoptic Survey Telescope\\
\url{http://www.lsst.org/lsst/}} \citep[e.g.][]{Gezari12}
and
\textit{eROSITA}\footnote{
extended ROentgen Survey with an Imaging Telescope Array\\
\url{http://http://www.mpe.mpg.de/eROSITA}}
\citep[e.g.][]{Khabibullin+13}.

This paper is organized as follows.
In \S\ref{sec:geometry}, I review the overall geometry of
the accretion flow onto the SMBH binary, namely
the formation of a central cavity
and the periodic leakage of discrete streams into this cavity.
I develop a heuristic model for the accretion of the stream in \S\ref{sec:accretion},
by estimating the thermal properties of the stream
after it is shock-heated and circularized (\S\ref{subsec:viscpres}),
and treating the subsequent gas accretion using
an idealized viscosity prescription (\S\ref{subsec:viscspr}).
I calculate mock slim-disc SEDs and light curves for two specific examples:
a $10^7~\Msol$ SMBH binary (\S\ref{subsec:energy}),
and a $10^9~\Msol$ binary whose orbital parameters
are motivated by the blazar OJ 287 (\S\ref{subsec:OJ287}).
In \S\ref{sec:pop}, I provide rough estimates
for the fraction of galaxies that could host SMBH binaries
producing quasiperiodic transient flares.
Observational implications, including prospects for detecting
the flares and distinguishing them from TDEs are discussed in \S\ref{sec:discussion}.
I conclude by summarizing the findings in \S\ref{sec:concl}.

\section{Accretion Flow Geometry}
\label{sec:geometry}

Let us consider a prograde accretion disc around a SMBH binary
of mass\footnote{
Throughout this work, I use $M$ to denote the mass
of a single SMBH, $M_{\bullet\bullet}$ for the total
mass of a SMBH binary, and $M_{\bullet}$ for the
mass of either member of a binary.
}
 $M_{\bullet\bullet}$, semimajor axis $a$ and orbital period $P$.
The Lindblad resonance clears the gas around the binary's orbit
and inhibits the infall of gas inward of a radius $\approx 2a$,
excavating an annular gap
(\citealt{ArtLub94}; but such a gap may close for low-mass binaries---see \citealt{Kocsis+12I, Kocsis+12II}).
For binaries with comparable BH masses,
the gas can cross inward of $\approx 2a$ only once per orbit
on nearly radial trajectories,
at a somewhat suppressed time-averaged rate---typically $\sim 10-100\%$
of the accretion rate at larger radii
(e.g. \citealt{Ochi+05}, \citealt{MM08}, \citealt{Cuadra+09}, \citealt{D'Orazio+12}).
It is convenient to think of the accretion flow in terms
of two dynamically distinct components separated by the gap\footnote{
These components are not completely independent of each other.
The gas that leaks into the cavity
can return to and interact with the
disc-cavity boundary \citep{Shi+12, Roedig+12, D'Orazio+12}.}:
an outer, geometrically thin and optically thick circumbinary disc
at $R\ga 2a$, and an inner, low-density region at $R\ltsim 2a$.
In this section, I present a rough physical description of each component.

\subsection{The outer disc}
\label{subsec:discs}

\subsubsection{Accretion disc around a single SMBH}
\label{subsubsec:disc1}
To best illustrate the effect of the central circumbinary cavity
on the observational appearance of the accretion disc,
I will briefly review the standard theory
for the spectral emission of a thin accretion disc
around a single SMBH.

Under the simple assumption that disc 
is a superposition of annuli that emit as isotropic graybodies
with a radial temperature profile $T_{\rm p}(R)$ for the thermalization photosphere,
the emitted SED $L_\nu$ is given by
\begin{align}
L_\nu&=2\int 2\pi R~F_\nu (T_{\rm p})~dR,\qquad\\
{\rm where}\qquad F_\nu (T_{\rm p})&=\pi \frac{2\epsilon_{\nu}}{1+\epsilon_{\nu}} B_\nu(T_{\rm p})
\label{eq:Lnu}
\end{align}
is the emergent flux per disc face,
$\epsilon_\nu$ is the ratio of the absorption opacity to the total opacity
\citep[see, e.g.,][]{Blaes04}, and $B_\nu$ is the Planck function.

The photospheric temperature profile  $T_{\rm p}(R)$  is given by
\beq
\Xi(R,T_{\rm p})~\sigSB ~T_{\rm p}^4(R)=\frac{3}{8\pi}
\left(1-\sqrt{\frac{R_{\rm ISCO}}{R}}\right)
\dot{M}\Omega_\K^2.
\label{eq:graybody}
\eeq
Above, $R_{\rm ISCO}$ is the 
radius of innermost stable circular orbit,
and the dimensionless factor $\Xi(R,T_{\rm p})$
quantifies the deviation of the emission from a blackbody  \citep{TM10};
at low temperatures (large disc radii), $\Xi\approx 1$.
The Keplerian angular velocity is denoted as $\Omega_{\K}$,
and $\dot{M}$ is the mass accretion rate,
which I will parameterize as
\beq
\dot{M}=\dot{m}\frac{L_{\rm Edd}}{\epsilon ~c^2},
\eeq
with $\dot{m}$ a dimensionless parameter
quantifying $\dot{M}$ in terms of the Eddington luminosity
$L_{\rm Edd}(M)=1.25\times 10^{38} ~(M/\Msol) ~\erg~\s^{-1}$
and a radiative efficiency $\epsilon$,
for which  I adopt $\epsilon=0.1$ throughout this paper.

The SED $\nu L_\nu$ peaks at a frequency 
\begin{align}
\nu_{\rm peak}^{\rm (single)}>
1.2\times 10^{16} 
&
\left(\frac{M}{10^7~\Msol}\right)^{-1/4}
\dot{m}^{1/4}
\nonumber\\
 \times &
\left(\frac{R_{\rm ISCO}}{6~GM/c^2}\right)^{-3/4}
\Hz,
\label{eq:nupeak1}
\end{align}
which is in the far UV or soft X-ray for SMBHs
accreting at a significant fraction of Eddington.
In equation \ref{eq:nupeak1}, the inequality is due to
graybody spectra having higher-frequency peaks than blackbodies;
the inequality symbol ``$>$''
can be replaced by ``$\approx$'' for blackbody SEDs.

The key point to take away from the above equations is that
the innermost regions of the accretion disc
are responsible for both the vast majority of the total emitted power,
as well as for the highest-energy thermal photons.

\subsubsection{Circumbinary disc truncated by tidal torques}
\label{subsubsec:disc2}
To a first approximation, the circumbinary disc can be thought
of as a standard AGN disc that is simply truncated by the cavity.
There are correcting factors---such as deformation of the surface density
profile outside the cavity \citep{SyerClarke95, IPP99, Kocsis+12II},
non-Keplerian potential variations
and pressure gradients at the disc-cavity boundary \citep{MM08}
and tidal heating of the edge of the disc \citep{Lodato+09b}---but
these do not significantly affect the theoretical exercise below,
and I will neglect them for its purpose.

The cavity size is expected to be much larger than $R_{\rm ISCO}$.
This means that circumbinary accretion discs are
``missing'' the innermost regions that produce
high-energy thermal photons, and will thus be
intrinsically dim at optical, UV and soft X-ray frequencies
compared to a disc around a single SMBH with the same mass.
The wider the binary, the larger the cavity and greater
the decrement in high-energy photons and luminous output.

In general, the size of the cavity $R_{\rm cav}$ with
respect to the binary's orbital separation $a$
is thought to lie in one of two distinct regimes, depending
on the balance between the timescale on which the binary's
separation decays, $t_{\rm res}\equiv | a/(da/dt)|$,
and the viscous diffusion timescale of the gas just
outside the cavity, $t_{\nu}$.
For binaries decaying more gradually than the disc's viscous diffusion timescale---i.e.,
if $t_{\rm res}>t_\nu$---the cavity is located at the resonance
radius $R\sim 2a$.
For very compact binaries, gravitational wave emission
causes the orbital separation to decay in a runaway fashion
($da/dt\propto a^{-3}$; \citealt{Peters64});
at a separation $a_{\rm dec}$ where $t_{\rm res}\sim t_\nu$, 
the binary ``decouples'' from the circumbinary disc,
rapidly decaying and merging before
the gas outside the cavity can significantly evolve \citep{MP05, TM10}.
Although some of the gas may be able to follow the binary
during this brief stage and produce some
high-frequency emission \citep{TMH12},
a strong density peak remains at the location $\sim 2a_{\rm dec}$.
Consequently, the peak of the SED of the circumbinary disc
does not evolve significantly between decoupling and merger,
and the SED brightens and hardens in earnest only
after the binary has merged.

Thus, during the lifetime of the binary, the surface density of
the circumbinary disc peaks at a radius $\sim \max(2a,2a_{\rm dec})$.
The peak frequency of the SED depends on the binary separation
(or equivalently, orbital period) until the binary decouples from the disc,
as
\begin{align}
\nu_{\rm peak}^{\rm (binary)}(a\ge a_{\rm dec})\sim 10^{14}
&
\left(\frac{M_{\bullet\bullet}}{10^7~\Msol}\right)^{1/4}
\dot{m}^{1/4}\nonumber\\
\times & \left(\frac{P}{10~\yr}\right)^{-1/2}
\Hz.
\label{eq:nupeak2a}
\end{align}
Between decoupling and merger, $\nu_{\rm peak}$
 is frozen at the value corresponding to $a\sim a_{\rm dec}$.
Although the exact decoupling separation depends
the properties of the binary and the disc,
several studies estimate $a_{\rm dec}\sim 100 ~GM/c^2$
for a wide range of parameters \citep{MP05, TM10, HKM09}.
Thus, 
\begin{align}
\nu_{\rm peak}^{\rm (binary)}(a\le a_{\rm dec})\sim
1.7\times &10^{15} 
\left(\frac{M_{\bullet\bullet}}{10^7~\Msol}\right)^{-1/4}
\dot{m}^{1/4}
\nonumber\\
 \times &
\left(\frac{a_{\rm dec}}{100~GM_{\bullet\bullet}/c^2}\right)^{-3/4}
\Hz.
\label{eq:nupeak2b}
\end{align}
Note that in contrast to discs around single SMBHs,
in circumbinary discs the deviation from blackbody
is small ($\Xi\approx 1$) and the thermal SED
has virtually zero dependence on the spin of either SMBH.

\begin{figure}
\centering
\epsfig{file=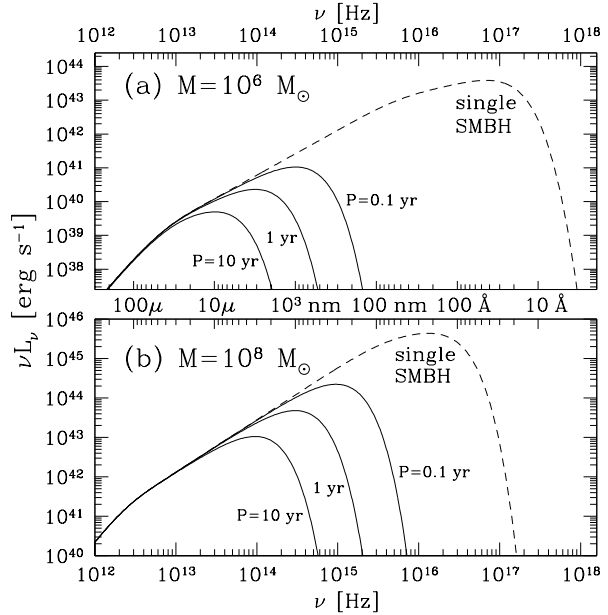, width=3.25in}
\caption{Model accretion disc SEDs with $\dot{m}=1$
and masses $M=10^8~\Msol$ (top panel) and $M=10^6~\Msol$ (bottom)
for the central engine.
In each panel, the dashed-curve SED that extends to high frequencies
is for a graybody disc around a single SMBH,
extending inward to $R_{\rm ISCO}=3~ GM/c^2$.
The less energetic spectra in solid lines correspond to discs around
binary systems with the same total mass $M$ and accretion rate,
but assumed to be truncated by a circumbinary cavity inside $R=2a$
for binary orbital periods $P=0.1~\yr$, $1~\yr$ and $10~\yr$.
Top and bottom panels show the cases for $M=10^6~\Msol$
$M=10^8~\Msol$, respectively.
The anomalously dim UV and X-ray emission, as well
as the unusual optical colors,
may cause some accreting SMBH binaries to elude AGN surveys.
}
\label{fig:outdisc}
\end{figure}

 In Figure \ref{fig:outdisc}, I have plotted model graybody spectra
for thin accretion discs around a single SMBH of mass $M$ extending to $R_{\rm ISCO}=3~GM/c^2$
(dashed lines) and binary SMBHs with the same total mass $M_{\bullet\bullet}=M$
and orbital periods $P=0.1~\yr$, $1~\yr$ and $10~\yr$,
truncated inside the corresponding radii $R=2a$ (solid lines).
Note that $P=0.1~ \yr$ coincides with decoupling in the $10^8~\Msol$ case.
The top and bottom panels show the cases for  $M=10^6 ~\Msol$ and $10^8 ~\Msol$, respectively.
All curves assume the temperature profile given in equation \ref{eq:graybody}
and $\dot{m}=1$.

The most prominent consequence of the central cavity
is that for a wide range of binary separations,
a circumbinary accretion disc is dim at UV frequencies,
unlike typical luminous AGN (see \citealt{TMH12}, \citealt{GultekinMiller12}).
If one assumes that the circumbinary disc does not extend
to outer radii where it is Toomre-unstable against gravitational fragmentation,
then accreting SMBH binaries would be most likely
to be detected with periods of years or decades 
(e.g. \citealt{HKM09}); however, the disc could be stabilized
at large radii by gravitational turbulence or additional heat sources,
(e.g. \citealt{Goodman03}).
For binaries with these periods, the cutoff in the disc SED
will occur at UV, optical or even infrared wavelengths.
Accreting SMBH binaries may therefore be missed by
AGN detection methods that rely on UV and optical
luminosities or colors.
If the hard X-rays in AGN are also produced in the central regions
of an accretion disc \citep[e.g.][]{Done+12},
then some binaries could also elude X-ray AGN surveys.

It is worth noting that the circumbinary disc is not
Eddington-limited in the usual sense.
Because the cavity reduces the luminous power
of the disc compared to single-SMBH systems,
the mass supply rate $\dot{m}$
of the circumbinary disc can be much higher than unity
(or equivalently, the surace density $\Sigma$ can be much higher than
in discs around SMBHs of the same total mass)
without the bolometric luminosity exceeding $L_{\rm Edd}$.

 There is some parameter degeneracy between the binary and single SMBH
values for $\nu_{\rm peak}$, so that circumbinary discs of
low-mass SMBH binaries can have SEDs that are similar in appearance
to very massive single SMBHs with low accretion rates---e.g.,
a $M_{\bullet\bullet}\sim10^7~\Msol$, $\dot{m}=1$
binary near decoupling has $\nu_{\rm peak}$ and $\nu L_\nu$
similar to that of a $M\sim 10^9~\Msol$, $\dot{m}=0.01$ single SMBH.
Such degeneracies may be broken through independent
determinations of the engine mass, or if either the gravitational-wave emission
from the merger or the subsequent UV/X-ray brightening of the SED \citep{MP05, THM10}
is observed.

The above contrast between discs around
binary and single SMBHs is a crude one, motivated by
the most basic implementation of the theory of geometrically
thin, optically thick accretion discs. In reality, observed AGN SEDs
can deviate greatly from the simple picture painted above.
Obscuration and dust-reddening of optical and UV photons can impede
simple comparisons motivated by the analytic theory,
especially if SMBH binaries preferentially reside in recently merged
and heavily enshrouded galaxy environments.

\subsection{Central cavity and leaked streams}
\label{subsec:streams}

The SEDs presented in Figure \ref{fig:outdisc} represent only a part of the total emission.
As stated above, the outer disc leaks gas periodically
into the cavity, and to predict the thermal emission from the whole system
we must understand what happens to this leaked gas.
Numerical simulations show that the leakage takes
place in discrete streams
that enter the cavity once per binary period on nearly radial free-fall trajectories
and preferentially accretes onto the secondary
\citep{ArtLub96, Cuadra+09}.

How far from either SMBH is the stream pericenter?
In a simulation of accretion onto a SMBH binary with a $1:3$ mass ratio,
\cite{Cuadra+09} found the accretion across a radius of $0.1a$ from either SMBH
to be highly punctuated and periodic, indicating that pericenter lies
inside this radius.
\cite{Sesana+12} resolve, in their simulations of a binary with a $0.35:1$ mass ratio,
small accretion discs with radii $\sim 0.005a$ around both SMBHs,
suggesting a similar upper limit to the pericenter radius.
(That is, the inner discs would be larger if the pericenters of the streams
were farther from the SMBHs.
Note that $0.1a$ is smaller than the canonical Hill stability radius,
$R_{\rm Hill}\ltsim a(1-e)[M_{\bullet}/(3M_{\bullet\bullet})]^{1/3}$,
where $M_{\bullet}$ is the mass of either SMBH
and $M_{\bullet\bullet}$ is the total binary mass.)
Similarly, the simulations of \cite{Hayasaki+08} for a $1:1$ system---under
the assumption that the streams cool efficiently
and the entire accretion flow onto the binary is isothermal---form
small, dense discs with radii $\sim 0.1a$.

If the streams accrete on timescales longer than the binary period,
then over time the leaked gas will form persistent, small, optically thick
accretion discs around one or both individual SMBHs \citep[as in][]{Hayasaki+08}.
In this scenario, the inner disc(s) would emit in the UV,
but more weakly than in single-BH AGN due to
the lower gas content \citep{TMH12,GultekinMiller12}.
The binary could also exhibit periodic optical/UV flares as
the incoming periodic streams interact
with the pre-existing inner disc(s) \citep{TMH12}
or the cavity wall \citep{Shi+12, Roedig+12, D'Orazio+12}.

An alternate possibility, which I explore further in this manuscript,
is that the leaked gas is consumed by one or both SMBH(s)
on a timescale faster than the binary's orbital period.
In this case, no persistent discs are present inside the cavity,
and the streams instead fuel transient accretion flows that
produce optical/UV/X-ray flares once per binary orbit.
In between streams, the accretion flow would exhibit
low intrinsic UV flux and unusual spectral shape
as discussed in \S\ref{subsec:discs},
and the system may appear as an odd AGN
or even masquerade as a quiescent nucleus.
As I discuss further in the next section, a
key motivation for postulating that the streams accrete
rapidly is the large amount of gravitational energy
that they have at pericenter, due to the streams being
marginally gravitationally bound to the binary.

The low-density cavity would be optically thin, and the diffuse gas
there is unable to cool efficiently.
Thus, in between the periodic leakage of the streams,
the central cavity should resemble the
underluminous, ADAF (advection dominated accretion flow) state 
\citep{Ichimaru77, NY94, NY95a}.
It is interesting to note that the conditions at
the boundary between the gap and the circumbinary disc---such as steep gradients in temperature, pressure
and radial velocity, as well as super-Keplerian orbital velocities
near the boundary---satisfy the theoretical requirements for consistent solutions
with an ``outer disc, inner ADAF'' configuration \citep{Honma96, Liu+99, ManmotoKato00}.
Despite producing virtually zero emission in the optical to soft-X-ray frequencies,
the optically thin accretion flow inside the cavity
may help generate outflows \citep{NY95a, BB99, Meier01, TM06, NM08},
which may be observable as radio sources or contribute to blazar emission (e.g. \citealt{FB99}, \citealt{Cao02}).

\section{Modeling the periodic flares}
\label{sec:accretion}

 I now turn to modeling the accretion of the leaked streams
 onto the individual SMBH(s), and the resulting emission.
I divide the accretion onto the binary SMBH in terms of
three distinct components, as follows.
\begin{enumerate}
\item \emph{Circumbinary disc.}
The  disc has a cavity maintained at $R\sim 2a$,
and has a surface density profile that corresponds to
a mass supply rate $\dot{m}_{\rm CB}\equiv \dot{m}(R\gg 2a)$. The thermal SED of the disc
is dim and soft compared to discs around a single SMBH
with comparable mass and accretion rate.
\item \emph{Leaked streams.}
Once per period, gas leaks into the cavity in the form
of elongated streams on nearly radial orbits.
Simulations indicate that the pericenters of the streams
occur inside $\ltsim 0.1a$. The amount of gas in the streams is equal
to or less than $P\times \dot{M}_{\rm CB}$.
The streams shock-heat near pericenter,
and their bulk orbital energy is dissipated as heat.
Some fraction of the gas circularizes its orbit
around one or both SMBHs and begin
to accrete through viscous diffusion,
with the rest being flung back out toward the circumbinary
disc or ejected as an outflow.
\item \emph{Post-shock accretion flow.}
The resulting hot, bound, post-shock gas fuels a transient, luminous
flare (or two flares) as it accretes onto either SMBH,
until it is depleted and becomes optically thin.
Then, the system is soft and dim again, until the next pair of streams
enter the cavity. \end{enumerate}

 \subsection{Viscosity Prescription} 
\label{subsec:viscpres}

 As the stream falls on a marginally bound orbit into the cavity,
it becomes tidally elongated.
Upon passing the point
of nearest approach to a SMBH, the leading end will collide with the rest
of the stream, shock-heating the gas;
additional heating may occur due to tidal compression,
as well as the enhanced velocity gradients and viscous shear at pericenter.
The expected result is that a moderate fraction
of the stream's  bulk kinetic energy is converted to radiation
 (see \citealt{Kim+99}, \citealt{SQ09} for similar calculations in the context of TDEs).

 Upon entering pericenter, the stream has a kinetic energy density
$\frac{1}{2} \sqrt{1-e_{\rm stream}^2}\rho v_\K^2$,
where $e_{\rm stream}\sim 1$ is the eccentricity of the stream
with respect to the SMBH to which it makes the closest passage,
$\rho$ is the gas density and $v_\K$ is the local circular Keplerian velocity.
The amount of energy that is converted to heat for whatever
portion of the stream that becomes (nearly) circularized
is thus $\sim\frac{1}{2}\left(1-\sqrt{1-e_{\rm stream}^2}\right)\rho v_\K^2$.
Radiation pressure dominates over gas pressure in the post-shock gas,
and the temperature $T_{\rm shock}$ is related to the 
local gravitational potential as
 \begin{align}
\frac{1}{2}\left(1-\sqrt{1-e_{\rm stream}^2}\right)\rho v_\K^2
&\sim\arad T_{\rm shock}^4,\qquad {\rm or}
\label{eq:shock1}\\
 \frac{1}{2}\rho v_\K^2 &\sim  3\rho \cs^2,
\label{eq:shock2}
\end{align}
where $\cs$ is the isothermal sound speed.

As alluded to in the previous section, the amount of orbital energy
that is available to shock-heat the accretion flow
 is very large; for the same depth of the gravitational potential,
the post-shock pressure in the gas is higher than
that a steady-state canonical $\alpha$-disc model \citep{SS73}
by a factor
\beq
\frac{ T_{\rm shock}^4}{ T_{\rm SS}^4}
\approx \frac{4}{27}\frac{\Xi}{\tau}\left(\alpha \frac{H}{R}\frac{\cs}{c}\right)^{-1},
\eeq
where  $\tau$ is the optical depth of the disc and
$H=\cs/\Omega_\K$ is the scale height.
For realistic disc properties, the above ratio is well in excess of unity.
Thus, if the viscous stress scales with the internal pressure of the fluid,
then the circularized post-shock gas will accrete much more rapidly
than gas in a steady-state disc.

What is the mass $M_{\rm flare}$ that is accreted in this fashion?
 As stated above, the time-averaged mass flux into the cavity is of order
$\sim 10-100\%$ of the mass accretion rate $\dot{M}_{\rm CB}\equiv \dot{M}(R\gg 2a)$
in the circumbinary disc, so the combined mass of the streams for each cycle is of order
$\sim (0.1-1)\times \dot{M}_{\rm CB}P$.
However, only a fraction of this mass will ultimately be bound to and
accreted by the SMBHs.     I therefore parameterize the mass of the post-shock gas that accretes onto
either SMBH in terms of this \textit{a priori} unknown fraction $f<1$: \begin{align}
M_{\rm flare}&=f ~\dot{M}_{\rm CB}~P\nonumber\\
&=0.2 ~\frac{f}{0.1}~\dot{m}_{\rm CB}~\frac{M_{\bullet\bullet}}{10^7~\Msol}~\frac{P}{10~\yr}~\Msol.
\end{align}
Whatever leaked mass is not accreted immediately may be
re-ejected into the circumbinary disc (see earlier footnote),
fall back onto either SMBH at a later time,
interact with and reprocess the photons produced
by the flare, contribute to the ADAF/corona within the cavity,
or ejected via thermally, radiatively or magnetically driven outflows.
While the unaccreted gas could have
observable features, the flare must be
the energetically dominant event, at least
in the UV and soft X-ray frequencies,
because it is the component that taps into the deepest parts of the
binary's potential well.

 Given the range of dynamic scales of the problem
and the number of relevant physical processes
involved---steep thermodynamic and density gradients,
asymmetries, the precise thermal and kinetic state of the post-shock bound gas,
the possible onset of viscous and thermal instabilities,
the importance of radiative forcing and photon trapping,
magnetohydrodynamic treatment of the viscosity mechanism,
to name a few---performing a fully consistent and detailed
calculation for the  evolution and accretion signature of the post-shock gas
 is a daunting task.
 Here, I instead construct a toy model of the problem,
employing an idealized viscosity treatment
which normalizes the viscosity through the canonical $\alpha$ prescription
and assumes the fluid elements have Keplerian circular orbits\footnote{
The results of the present analysis should not be significantly affected
if the bound gas forms an eccentric accretion disc;
see \cite{SyerClarke92,Lyubarskij+94,Ogilvie01}.
}.
This approach is admittedly crude;
it is meant only to serve as a proof-of-concept that
the post-shock accretion event can develop and decay rapidly
and produce a luminous outburst.
The toy treatment presented here
could also provide an analytic foundation within which future,
more detailed studies may be framed.

The post-shock gas that is bound to either SMBH will spread
as turbulent viscosity transports angular momentum and dissipates orbital energy.
 The kinematic viscosity $\nu$ is the product of the gas sound speed
and the size of the largest turbulent eddies. The latter cannot be
larger than the disc scale height $H=\cs/\Omega_\K$, so I parameterize
the viscosity in the familiar fashion,
$\nu=\alpha H \cs=\alpha \cs^2/\Omega_\K$,
with $\alpha<1$ \citep{SS73}.
 Using the relationship between the sound speed
and the local Keplerian circular velocity in the post-shock gas in equation \ref{eq:shock2}, I prescribe
\beq
\nu=\alpha\frac{\cs^2}{\Omega_{\K}}
\sim \frac{\alpha}{6}\frac{v_{\K}^2}{\Omega_{\K}}
=\frac{\alpha}{6}R^2 \Omega_{\K}.
\label{eq:visc}
\eeq
Empirical estimates of $\alpha$ in AGN typically give $\alpha \sim 0.01 - 0.1$,
and simulations of the magnetorotational instability suggest similar values \citep{Pessah+07}.
There are indications that outbursting systems have higher values, $\alpha\sim 0.3$
(e.g., \citealt{Dubus+01}, \citealt{Starling+04}, \citealt{King+07});
dissipations of supersonic shocks or other instabilities
could drive macroscopic turbulence with $\alpha\sim 1$
\citep[e.g.,][]{Carciofi+12}.
In magnetohydrodynamic simulations of circumbinary discs,
\cite{Shi+12} and \cite{Noble+12} find Reynolds and Maxwell
stresses equivalent to $\alpha >1 $ near the cavity edge.

The $\alpha$ prescription in equation \ref{eq:visc} should be
viewed as an approximate normalization of the magnitude of the viscous
stresses in the post-shock accretion flow.
In reality, the effective value of $\alpha$ is unlikely to be a constant
with respect to time or radius, owing to the inhomogeneities and
sharp thermodynamic gradients;
similarly, the radial viscosity profile may evolve to deviate
from the $\nu \propto R^{1/2}$ post-shock profile.
In addition, the post-shock accretion flow may be susceptible
to thermal or viscous instabilities \citep{LE74, SS76, Pringle76, Piran78, HKB09,Jiang+13}.
If the accretion flow in question is thermally unstable,
it may undergo runaway heating and be
accreted more rapidly than in the idealized calculation presented here.

 \subsection{Viscous Spreading} 
\label{subsec:viscspr}

Treating the post-shock, bound accretion flow as a superposition
of annuli centered around either SMBH,
I use the standard equation
for Keplerian, viscous thin discs \citep[e.g.,][]{Pringle81},
which gives the evolution of the vertically integrated density $\Sigma$:
\beq
2\pi R\frac{\dd}{\dd t}\Sigma (R,t)=\frac{\dd}{\dd R}\left[2 R^{1/2}\frac{\dd}{\dd R}\left(3\pi\nu\Sigma R^{1/2}\right)\right].
\label{eq:thindisc}
\eeq
The idealized viscosity form in equation \eqref{eq:visc},
$\nu \propto R^{1/2}$,
has well-documented exact solutions
for arbitrary initial profiles of $\Sigma$
\citep{Lust52, LP74, Pringle91, Tanaka11}
that I will employ throughout the rest of this paper.

 Initially, the post-shock accretion flow will be moderately geometrically thick,
with $H^2/R^2\sim \cs^2/v_{\K}^2 \sim 1/6$.
The pressure is dominated by radiation,
and the main source of opacity is electron scattering.

The viscous diffusion timescale of the gas is
\beq
t_{\nu}(R)=\frac{R}{v_R}=\frac{4}{\alpha}\left[1+2\frac{d\ln(\nu\Sigma)}{d\ln R}\right]^{-1}\Omega_{\K}^{-1}.
\eeq
For an initial mass distribution initially concentrated at a radius $R_0$,
\beq
\frac{t_\nu (R_0)}{P}
<
\frac{2}{\alpha^\prime \pi}\left(\frac{M_{\bullet\bullet}}{M_{\bullet}}\right)^{1/2}\left(\frac{R_0}{a}\right)^{3/2}.
\eeq
That is, the accretion flow will be depleted on a timescale
shorter than the binary orbital period \citep[cf.][]{Sesana+12}
if it is initially deposited at a radius
\begin{align}
R_0&\ltsim 0.3 \left(\frac{\alpha}{0.1}\right)^{2/3}\left(\frac{M_{\bullet}}{M_{\bullet\bullet}}\right)^{1/3}a\nonumber\\
&\sim 1.0\left(\frac{0.4}{1-e}\right)\left(\frac{\alpha}{0.1}\right)^{2/3}R_{\rm Hill}.
\label{eq:Rcomp}
\end{align}
Above, I set $e\sim 0.6$ as a reference value for the binary eccentricity
\citep{Roedig+11}. 
As discussed  in the previous section (\S\ref{subsec:streams}),
 inequality \ref{eq:Rcomp} is amply satisfied by the orbital trajectories of the leaked streams in
numerical studies.

       The regions where the accretion flow is optically thin
cannot be treated by the thin-disc equation \ref{eq:thindisc},
and will have negligible viscous transport
and radiative output compared to the optically thick regions.
I compute $\tau=\theta\kappa_{\rm es}\Sigma$,
where $\theta=0.2$ is a porosity factor \citep{Turner04},
and neglect the emission from the regions where $\tau<1$.
Because the accretion flow spreads so that the surface density
at large radii is lower than near the center \citep[see, e.g., ][]{Tanaka11},
the outer regions of the disc become optically thin first.

Let us now apply this model to a specific example.
I consider a binary of total mass $M_{\bullet\bullet}=10^7~\Msol$,
with a period of $P=30~\yr$ ($a=0.010~\pc$).
I set $\dot{m}_{\rm CB}=1$ and $f=0.05$, which fixes
the total mass in the bound, post-shock accretion flow at $M_{\rm flare}=0.33~\Msol$.
I assume accretion onto the secondary SMBH with $M_{\bullet}/M_{\bullet\bullet}=1/10$
and $R_{\rm ISCO}=3~GM_{\bullet}/c^2$.
I solve the viscous evolution of the post-shock annulus
for two different initial surface density distributions.
One is a $\delta$-function ring
initially located at a radius $R_{0}=0.05a\approx 10^4~ GM_{\bullet}/c^2$
from the SMBH.
The second initial profile is a torus with constant gas volume density
(i.e., $\Sigma \propto R^{-1}$)
that extends from $0.5~R_0$ to $1.52~R_0$.
The dimensions of the second profile were chosen
so that both initial conditions share the same values
for the total orbital angular momentum and gas mass.

\begin{figure}
\centering
\epsfig{file=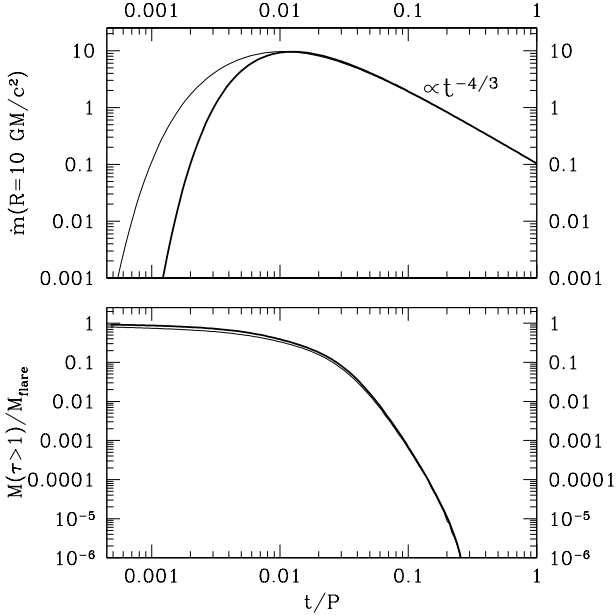, width=3.25in}
\caption{
{\it Top panel}: The mass flux $\dot{M}$ evaluated at $R=10~GM_{\bullet}/c^2$
as a function of time, for two initial density profiles:
a $\delta$ function centered at $R_0=0.05a$ (thick curves)
and a constant-density annulus extending from $0.5R_0$ to $1.5R_0$.
The total mass of the stream is initially $M_{\rm flare}=0.05 \dot{M}_{\rm crit}P$,
where  $\dot{M}_{\rm crit}$ is the critical accretion rate for the secondary mass.
{\it Bottom panel}: The fraction of the initial mass in the accretion flow
that is optically thick $\tau>1$, assuming electron-scattering opacity.
There are two (nearly overlapping)
curves for each of the two initial conditions.
}
\label{fig:mdot}
\end{figure}

In the top panel of Figure \ref{fig:mdot}, I plot the mass flux
\beq
\dot{M}(R,t)=3\pi\nu\Sigma\left[1+2\frac{d\ln(\nu\Sigma)}{d\ln R}\right]
\left(1-\sqrt{\frac{R_{\rm ISCO}}{R}}\right)^{-1}
\label{eq:mdot}
\eeq
evaluated at $R=10~GM_{\bullet}/c^2$,
for an individual accretion episode onto the secondary SMBH.
The evolution of the $\delta$-function ring is plotted with a thin curve,
and that of the constant-density torus is plotted with a thick curve.
For both initial profiles, the accretion rate peaks at nearly the same time,
and the subsequent decay behavior is virtually identical.
 Note that although the accretion rate reaches values above
the ``critical'' Eddington value (i.e. $\dot{m}>1$),
the luminosity produced by the accretion event
is capped at near $L_{\rm Edd}$ (see \S\ref{subsec:energy} below)
due to radiative advection.

For our viscosity prescription,
the late-time mass accretion rate scales as $\dot{M}\propto t^{-4/3}$,
\beq
\frac{d \ln \dot{M}}{d\ln t}\approx -1-\left(4-2\frac{d\ln\nu}{d\ln R}\right)^{-1}=-\frac{4}{3},
\eeq
due to the late-time behavior of the central surface density profile \citep{Tanaka11}.
This is quite similar to the canonical late-time behavior $\dot{M}\propto t^{-5/3}$
for TDEs \citep{Rees88, Phinney89},
which arises from the orbital energy distribution of the disrupted
stellar material.
For both the transient flares discussed in this paper (see \S\ref{subsec:energy} below)
and TDEs \citep{Lodato+09a,GuillochonRR13, ShenMatzner12},
the light-curve at any given frequency is not expected to follow
the same power-law decay as the mass accretion rate.
Thus, for either type of transient, it is unlikely that the
underlying mass accretion rate could be accurately deduced
from the observed emission.
In other words, it may not be possible to distinguish 
a cavity-accretion flare ($\dot{M}\propto t^{-4/3}$
in this idealized treatment) from a TDE ($\dot{M}\propto t^{-5/3}$)
solely from the decay in the light curve.

A key model assumption with respect to the light curve is that
the accretion flow forms promptly, i.e. on timescales shorter than $\sim 10\%$
of the binary orbital period.
The fact that the dynamical times are much shorter where the stream shocks ($R<0.1a$)
favors this hypothesis, but this should be confirmed by numerical simulations.
If the fueling accretion flow materializes on timescales longer than the fuel depletion timescale,
then the light curves would rise more slowly.
 Because a promptly rising light curve is a key smoking-gun
feature of TDEs, if cavity-accretion flares develop more gradually,
this could be used to distinguish the two types of transients.

In the bottom panel of Figure \ref{fig:mdot}, I plot the fraction
of the initial gas mass that is optically thick ($\tau>1$).
For fiducial parameter values, the entire stream becomes optically thin
on a timescale shorter than $P$, at which point the entire flow will have
transitioned into a radiatively inefficient and under-luminous state.
The amount of time it takes the gas to be depleted in this way
depends on the total amount of gas in the flow.
For very high accretion rates into the cavity, 
the post-shock flow may not be depleted before the next stream
arrives at pericenter;
in this case, persistent mini-discs would form around one or both SMBHs
and the flares would be less dramatic.
The exact fraction of time the system spends in quiescence (as an odd-colored
or underluminous AGN) depends on the amount of the leaked streams
that is consumed by the SMBH(s) (i.e. on $\dot{M}_{\rm CB}$
and the unknown parameter $f$), as well as the binary period $P$.

\subsection{Energetics and Emission}
\label{subsec:energy}
To model the emergent spectrum, I now estimate
the radial temperature profile of the disc.
I assume that the accreting gas is able to promptly dispose
of the viscously generated heat via graybody radiation and advection:
\beq
Q_\nu \sim Q_{\gamma}+Q_{\rm adv},
\label{eq:ebalance}
\eeq
where
\begin{align}
Q_{\nu}&=\frac{9}{4}\Sigma \nu \Omega^2,
\label{eq:Qnu}\\
Q_{\gamma}&=2\times \Xi \frac{4\sigSB T^4}{3\tau},\qquad {\rm and}
\label{eq:Qgam}\\
Q_{\rm adv}
&=-\frac{\dot{M}}{2\pi R^2}\cs^2\left(12\frac{d\ln T}{d\ln R}-4\frac{d\ln \rho}{d\ln R}\right)
\label{eq:Qadv}
\end{align}
represent viscous heating, radiative cooling
and advective cooling \citep{Abramowicz+88}, respectively.
Outer regions of the accretion flow (where the surface density is lower)
that are optically thin are assumed to be radiatively inefficient,
and excluded from the emergent flux calculation.

\begin{figure}
\centering
\epsfig{file=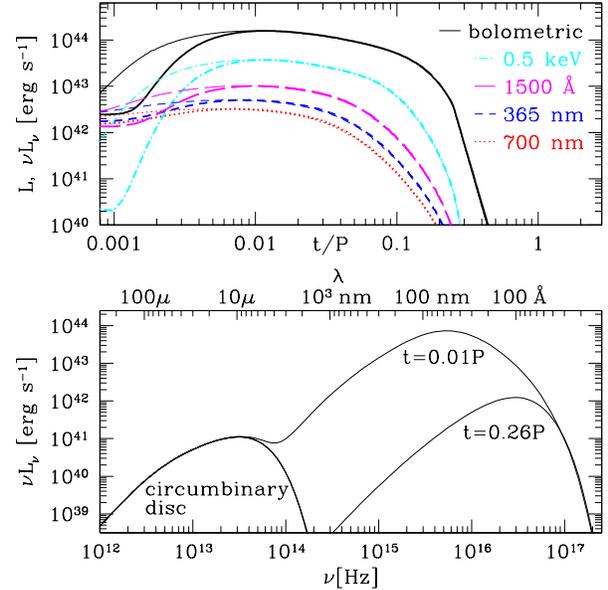, width=3.25in}
\caption{
Toy light curves (\textit{top panels}) and
spectrum snapshots (\textit{bottom panels})
for a $M_{\bullet\bullet}=10^7~\Msol$ binary with a $1:10$ mass ratio.
The binary has a period of $30~\yr$, which places the
peak emission frequency of the circumbinary disc
in the infrared. In this example, the flares peak in the UV and soft X-ray
frequencies and last several months before decaying rapidly
over several years. Neither the bolometric nor the monochromatic
light curves obey the $t^{-4/3}$ decay of the central accretion rate.
}
\label{fig:lcurves}
\end{figure}

In Figure \ref{fig:lcurves},
I present toy-model light curves, along with several snapshots of the emergent spectrum,
for the accretion event modeled earlier in Figure \ref{fig:mdot}.
The binary mass is $M_{\bullet\bullet}=10^7~\Msol$,
the period is $P=30~\yr$, and the accretion occurs onto the
secondary SMBH with $M_{\bullet}=10^6~\Msol$.
I have included for reference the emission from the
circumbinary disc, with $\dot{m}_{\rm CB}=1$, $\alpha=0.1$ and a cavity at $R<2a$.

The top panel shows the bolometric light curve
(black, solid curve)
along with several monochromatic ones:
R-band ($700~{\rm nm}$; red, dotted curve),
U-band ($365~{\rm nm}$; blue, short-dashed curve),
far UV ($1500~\AA$; magenta, long-dashed curve), and
0.5 keV (cyan, dash-dot curve).
As in Figure \ref{fig:mdot}, thick lines show the light curves for the $\delta$-function
initial density profile, and thin curves show those
for the constant-density profile.
Both light curves converge near their peaks,
occurring at roughly $t=0.01P$. The subsequent emission drops off steeply
as the gas is depleted and the outer accretion flow becomes
optically thin and radiatively inefficient.

None of the light curves follow a $L\propto t^{-4/3}$ power-law decay of the central
mass accretion rate (Figure \ref{fig:mdot});
as is the case with TDEs, it is unlikely that the underlying
accretion rate can be deduced from the observed light curves
(see \S\ref{subsec:viscspr}). 
The peak bolometric luminosity is roughly equal to the
Eddington luminosity of the secondary, even though the accretion
rate is supercritical by an order of magnitude.
 This is consistent with models of slim accretion discs
 (e.g. \citealt{Watarai+00}, \citealt{Ohsuga+05}, in which advection helps to stabilize
the disc against radiative self-destruction even when $\dot{m}\gg 1$.

In the bottom panel of Figure \ref{fig:lcurves},
I show several snapshots of the SED,
calculated as a graybody spectrum from the temperature profile
as in the circumbinary SEDs in Figure \ref{fig:outdisc}.
On the lower-left corner is the emission of the circumbinary disc,
which in this example peaks in the infrared and is optically dim.
This is the emission from the system most of the time, in between the flares.
The composite (flare plus circumbinary disc) spectrum
at the peak ($t=0.01P$) and just as the flare is fading ($t=0.26P$)
are also shown to demonstrate that, in principle, the
time evolution of the emission properties can be quite dramatic.
In this toy example, the binary 
spends $\approx 70-90\%$ of its time in quiescence,
with the flares decaying to less than $\sim 10\%$ of the peak
luminosity within $\sim 0.1 P$.
However, in addition to the viscous and thermal evolution of the accreting gas,
the duration of the flare is also sensitive to the raw amount of mass
deposited via the stream.

\subsection{An alternate binary model for OJ 287}
\label{subsec:OJ287}
 The BL Lacertae object OJ 287 is known to produce
pairs of optical outbursts every $11-12$ years.
The outbursts occur in pairs whose peaks are
separated by a nearly constant interval of
approximately 2 years.
While the flux amplification values from episode to episode,
the optical brightness is observed to rise by as much as 5 magnitudes
during a flare.

Beginning with \cite{Sillanpaa+88}, many studies have interpreted this system as
a SMBH binary.
In particular, \cite{LV96}, \cite{Valtonen+06} and \cite{Valtonen+08}
have modeled the flares as resulting from
a secondary SMBH impacting a circumprimary accretion disc.
In this disc-impact model, the secondary's orbit
is elliptical and inclined with respect to the accretion disc,
so that two impacts occur near pericenter.
\cite{Valtonen+12} take the primary and secondary masses to be
$1.8\times 10^{10}~\Msol$ and $1.2\times 10^8~\Msol$, respectively.

Here, I also interpret OJ 287 as an accreting binary with a period of 11.7 years
(9 years in the binary's reference frame, for redshift $z=0.3$)
but with two substantive differences from the model of Valtonen et al.
First, the binary's orbit is coplanar with a circumbinary accretion disc,
and the disc has a circumbinary cavity with a radius of $2a$.
Second, I use a total binary mass of $10^9~\Msol$, more consistent
with independent observational results than the $>10^{10}~\Msol$ value
in the binary model (e.g., \citealt{Xie+02}, \citealt{LiangLiu03}, \citealt{FanCao04}, \citealt{Gupta+12}).
The accretion rate in the circumbinary disc is taken to be
$\dot{m}_{\rm CB}=5$, by matching the optical-wavelength flux
in between outbursts with the circumbinary disc output.
(As noted in \S\ref{subsubsec:disc2}, because of the cavity
drastically reduces the total disc power,
the circumbinary disc can have $\dot{m}_{\rm CB}>1$ without being Eddington-limited.) 
I take the primary and secondary masses to be $8\times 10^8~\Msol$
and $2\times 10^8~\Msol$, respectively.
I suppose, arbitrarily, that they accrete $f=5\%$ of $\dot{m}_{\rm CB}$ between them,
and that the fraction consumed by each SMBH is inversely proportional to their mass ratio.
I take $\alpha=0.5$ and $R_0=0.05a$ for the post-shock accretion flows
(note the observational indications that $\alpha\sim 1$ for violent accretion events of this type; \S\ref{subsec:viscspr}),
and assume that the flares are offset by 2 years in the observer's frame,
based on the observed delay between the two peaks.

\begin{figure}
\centering
\epsfig{file=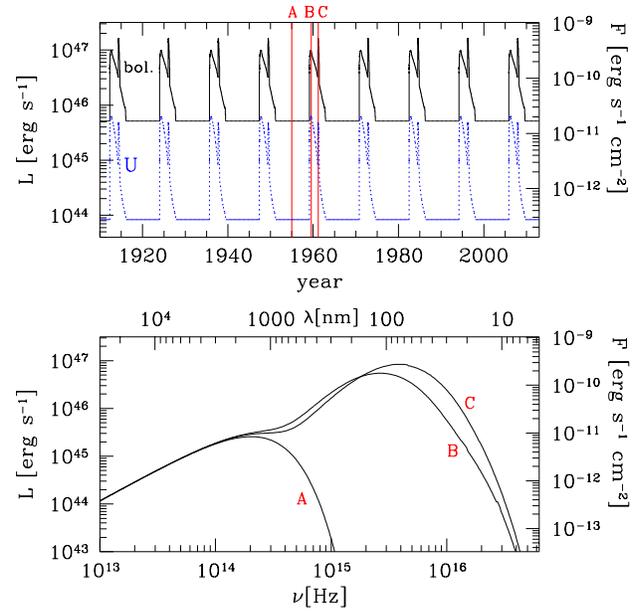, width=3.25in}
\caption{
A toy emission model for OJ 287.
The system is modeled as a binary with a total mass
$M_{\bullet\bullet}=10^9~\Msol$, a $1:4$ mass ratio,
a redshift of $z=0.3$ and an orbital period of $P=9~\yr$
($P_{\rm obs}=11.7~\yr$).
The binary's tidal torques are assumed to deposit a stream
near each SMBH, producing flares that are offset by $2~\yr$ (see text for details).
\textit{Top panel:} the bolometric (black)  and U-band (blue) light curves,
combining the emission from both flares and the circumbinary disc.
\textit{Bottom panel:} Snapshots of the SED
during quiescence (marked with a vertical line and the letter ``A'' in the light curve),
during the first flare (``B'') and during the second flare (``C''). 
}
\label{fig:OJ}
\end{figure}

I present the light curves and SED snapshots of this model in Figure \ref{fig:OJ}.
The observable flux is computed from the source emission
assuming a luminosity distance of $1600~{\rm Mpc}$ to correspond
to the source redshift.
The bolometric luminosity (black solid curves, top panel) fluctuates by an order of magnitude
between quiescence and flaring. The U-band monochromatic flux (blue dotted curves)
varies by two orders of magnitude.
SEDs during quiescence (``A'') and the peak of each flare (``B, C'') are shown in the bottom panel.
The first flare, due to accretion onto the secondary, lasts longer because
I have assumed that this SMBH accretes more gas than the primary.

Many different spectra and light curves can be produced by varying
the model parameters, but the production of two recurring flares
at near-Eddington fluxes is a generic feature for the cavity-accretion
flares discussed here.
Given the approximate nature of the model, I undertake no special
effort to obtain a detailed fit to the observed properties of OJ 287.
Rather, the goal of this exercise is merely to demonstrate that
this simple picture---periodic streaming of gas into the disc cavity and subsequent
shock-heating and rapid accretion onto the SMBHs---can
provide an explanation for the cadence,
duration and energetics of the optical flares.

With respect to OJ 287, the present model has several notable features.
First, the interpretation does not require a primary mass of
$\ga 10^{10}\sim\Msol$, which is over an order of magnitude
larger than several empirical estimates, as well as the $M_\bullet-\sigma$ relation.
Second, the cavity naturally explains the optical dimness of the system
in between outbursts. In fact, the steep dropoff in the SED at optical
frequencies arises simply from a $\dot{m}\gta 1$ disc around a $M\sim 10^9~\Msol$
binary being truncated at $R\sim 2a$.
However, this optical reddening could also be caused by dust extinction,
which could be enhanced due to the galaxy major merger that produced the SMBH binary \citep{Valtonen+12}.
Depending on the spins of the SMBHs, one or both flares
in each cycle may be radio-loud \cite{Valtaoja+00}.
Additionally, the shocks that trigger the formation
of the transient accretion flows
could explain the optical precursor flares
observed before the major outbursts \citep{Valtonen+06, Pihajoki+13}.

In the disc-impact model of \cite{Valtonen+06}, the flares
are caused by a precessing secondary impacting a circumprimary
disc that is stationary in space; the timing predictions
are precise and rigid, being based on post-Newtonian parameter fits
for the orbital decay and precession rates.
In contrast, in the cavity-accretion flare model I discuss here,
some scatter in the flare timing is expected due to
the intrinsic variability in the dynamics of the streams.
Future observations could discriminate the two models by
confirming or falsifying the binary orbit solution of the disc-impact model.
If future flares show a timing scatter that
cannot be fit by a precessing secondary impacting a fixed disc,
this would favor quasi-periodic cavity-accretion
as the mechanism behind the OJ 287 flares.

\section{Population of flaring SMBH binaries}
\label{sec:pop}

If cavity-accretion flares do indeed occur in nature,
how frequent are they?
The formation and dynamical evolution of sub-parsec SMBH binaries
remain active topics of research
in which guiding empirical clues are scarce \citep[see, e.g.,][]{ColpiDotti11}.
However, we have some concrete theoretical expectations,
such as the fact that a galaxy merger precedes SMBH binary formation,
as well as a robust theoretical foundation for the hierarchical
growth of galaxies.
Below, I will make use of this theoretical knowledge
to make educated estimates for the global event rate
of cavity-accretion flares.

\cite{Fakhouri+10} found that the mean merger rates
of dark matter halos 
in the  Millennium Simulations \citep{Springel+05,BoylanKolchin+09b}
can be described by a simple analytic 
fitting formula.
For mergers with halo mass ratios between unity and $30:1$,
the merger rate (per halo per unit time) evaluates to
\beq
\Gamma_{\rm merger}\approx 2.1
\left(\frac{M_{\rm halo}}{10^{12}~\Msol}\right)^{0.13} ~(1+z)^{0.099}
\frac{dz}{dt},
\label{eq:halomerge}
\eeq
for a given halo mass $M_{\rm halo}$ and redshift $z$.
I compute $dz/dt$ in the standard $\Lambda$CDM cosmology
with $h=0.70$, $\Omega_{\Lambda}=0.72$ and $\Omega_{\rm m}=0.28$
\citep{Hinshaw+12}.

By relating the stellar velocity dispersion $\sigma_*$ of galaxies to the
dark matter halo mass, one can attempt to convert the observed
$M-\sigma_*$ relation \citep{FerrareseMerritt00, Gebhardt+00, Gultekin+09, Graham+11}
to a relation between $M$ and the host halo mass $M_{\rm halo}$
(\citealt{Ferrarese02}; cf. \citealt{WL03, Baes+03, Bandara+09}).
Following \cite{GrahamScott13}, we adopt a broken power-law
\beq
\frac{M}{10^8 ~\Msol}\sim 
3 \left(\frac{M_{\rm halo}}{7\times 10^{12}~\Msol}\right)^{A_M}
(1+z)^{3/2},
\label{eq:mbhmhalo}
\eeq
where the slope $A_M$ equals $5/3$ if $M_{\rm halo}<7\times 10^{12}~\Msol$
(or $M<3\times 10^8~\Msol$)
and $6/5$ at higher masses.
Although this relation is known to have a large intrinsic scatter,
it serves to provide an order-of-magnitude estimate.
Combining equations \eqref{eq:halomerge}
and \eqref{eq:mbhmhalo}, one can estimate the rate
at which two SMBHs with combined mass $M_{\bullet\bullet}$
end up in a single halo.

Some fraction of such pairs will form binaries embedded in circumbinary
discs. There are (semi-) analytic formulations of how such binaries
evolve toward merger \citep{SyerClarke95, IPP99, LiuShap10, HKM09}.
In the simplest description, the evolution can be written
\beq
\frac{da}{dt}=\dot{a}_{\rm GW} + \dot{a}_{\rm disc}.
\eeq
At large separations, the binary evolves by depositing its
orbital angular momentum into the circumbinary accretion disc
($\dot{a}_{\rm disc}$), and at close separations by emitting gravitational
waves ($\dot{a}_{\rm GW}$).
For the former, we are interested in secondary type-II migration,
where the binary opens a central cavity in the disc.
I take the simple prescription given by \cite{SyerClarke95},
\beq
\dot{a}_{\rm disc}=-\frac{a}{t_\nu}
\left[\frac{2\dot{m}(1+q)^2}{q} \frac{t_\nu}{t_{\rm Sal}}\right]^k,
\label{eq:adotdisc}
\eeq
where $q\le 1$ is the secondary-to-primary mass ratio,
$t_\nu=2R^2/(3\nu)$ is the viscous diffusion timescale evaluated at the cavity edge $R=2a$,
and the dimensionless quantity $k\approx 0.4$ depends on the  
surface density and viscosity profiles of the disc.
I take $\nu$ to be given by a standard Shakura-Sunyaev disc
with $\alpha=0.1$ and $\dot{m}_{\rm CB}=0.3$ \citep[typical for luminous AGN; e.g.][]{Kollmeier+06}.
I assume $q=0.1$ as a typical SMBH mass ratio in such binaries.
The effects on the above prescriptions for varying the viscosity prescription
and various system parameters (e.g., $q$, $\dot{m}$, $\alpha$)
can be found in \cite{HKM09};
given the approximate nature of this exercise, I will not explore
this parameter space.
Note that \cite{Kocsis+12II} find
that the disc-driven migration of the binary could be
significantly slower than the expression in equation \eqref{eq:adotdisc};
the actual lifetimes of SMBH binaries with circumbinary discs
 could be longer than estimated here.
For the gravitational wave-driven evolution, I take
\beq
\frac{\dot{a}_{\rm GW}}{GM/c^2}
=
\frac{64}{5}\frac{c^3}{GM_{\bullet\bullet}}
\frac{q}{(1+q)^2}\left(\frac{a}{GM/c^2}\right)^{-3}
\eeq
from \cite{Peters64}, assuming circular orbits for simplicity.

The most liberal assumption possible is that
once SMBH pairs are formed after a galaxy merger,
all of them evolve toward coalescence while embedded in circumbinary discs.
Thus, an upper limit for the average flaring rate per galaxy $\langle\Gamma_{\rm flare}\rangle$,
\textit{averaged over all galaxies,} can be 
estimated by counting the total number of flares (orbits)
a binary makes while embedded in a disc,
and then multiplying by the universal formation rate 
of binaries (i.e., the halo major merger rate):
 \beq
\langle\Gamma_{\rm flare}\rangle\sim \langle P^{-1}\rangle
< \Gamma_{\rm merge}(M_{\bullet\bullet}, z)
\int \left(\frac{da}{dt}\right)^{-1} \frac{1}{P} ~da.
\label{eq:gamflare} 
\eeq
I've assumed one flare per binary orbit above, but it is trivial
to generalize to two flares.
Similarly, the maximum fraction of galaxies that host
a flaring SMBH binary between observed orbital periods
$P_1$ and $P_2$ can be estimated by
 multiplying the
total time a binary spends evolving between those periods
with the formation rate of binaries: \beq
f_{\bullet\bullet}(P_{\rm obs})
< \Gamma_{\rm merge}(M_{\bullet\bullet}, z)
\int_{a(P_2,z)}^{a(P_1,z)} \left(\frac{da}{dt}\right)^{-1} ~da.
\label{eq:fflare} 
\eeq
Note that above, $f_{\bullet\bullet}$ is the fraction of objects
that exhibit flares, not the fraction of time SMBHs spend flaring as
opposed to quiescence.
The $z$-dependence of the above expressions assumes
that the timescale for the formation of a disc-embedded
binary takes place on timescales much shorter than
a Hubble time.

I conservatively choose the upper limit of integration in equations
\eqref{eq:gamflare} and \eqref{eq:fflare} to be
$a_Q\equiv R(Q=1)/2$, i.e. that the disc does not extend
to regions where it is canonically Toomre-unstable (\S\ref{subsec:discs});
if the disc is stabilized by other mechanisms such
as gravitational turbulence, the number of flaring objects and flares in the sky
would be higher than estimated here.
I limit the integration to a minimum orbital period
in the observer rest frame $P_{\rm obs}=(1+z)P$
of $1$ month, to focus on flares whose quasiperiodically recurring
nature may be challenging to detect.
(Objects whose flares recur on shorter timescales
would be identified as such shortly after being detected,
in the course of standard monitoring of a TDE-like transient.)
I do not count binaries that have already decoupled from their
circumbinary discs, as such systems are expected to
produce weaker flares (if marginally decoupled) or no flares at all.
If some circumbinary gas is able to follow the binary in
the gravitational wave-driven regime \citep{TMH12},
flaring SMBH binaries may also be detectable in gravitational waves by pulsar timing arrays.
 Note that we can immediately place an order-of-magnitude estimate
that $f_{\bullet\bullet}\sim 10^{-3}$,
since we know $\Gamma_{\rm merge}\sim {\rm a~few}\times \Gyr^{-1}$
and that the timescale for a SMBH binary to evolve from
$a_Q$ to coalescence is $\sim t_\nu \sim {\rm Myr}$.

\begin{figure}
\centering
\epsfig{file=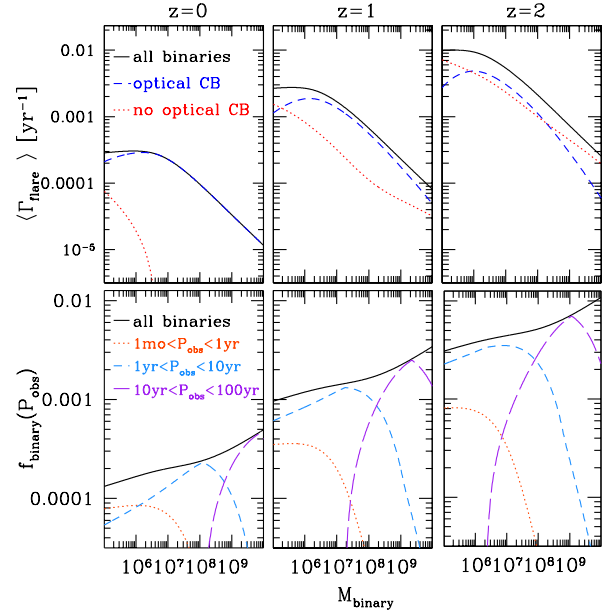, width=3.25in}
\caption{
\textit{Top panels:} Estimated upper limits of the flaring rate per galaxy,
across all galaxies, in the source rest frame. The black, solid curves
show the rate for all binaries with separations between
the value a gravitationally stable circumbinary disc can exist
and the value where the observed binary period is 1 month.
The blue, dashed curves show the rate only from binaries
whose circumbinary discs are visible as optical AGN,
and the red, dotted curves show the sources whose
circumbinary discs are optically dim.
These upper limits show that the global flaring rate could, in principle,
exceed the rate of stellar TDEs.
\textit{Bottom panels:}
Upper limits for the fraction of all galaxies that host
flaring SMBHs.``All binaries'' refers to the same range of semimajor
axes as in the top panels,
and the colored lines show the fractions for those binaries
in the denoted ranges of orbital periods in the observer's rest-frame.
For both the top and bottom panels, the redshift of the population increases
from left to right as $z=0,~1,~2$.
}
\label{fig:pop}
\end{figure}

I plot these upper limits for
$\Gamma_{\rm flare}$ and $f_{\bullet\bullet}$
in Figure \ref{fig:pop}, as a function of $M_{\bullet\bullet}$
and for several different subsets of the binary SMBH population.
From left to right, the vertical pairs of panels show
binary populations at $z=0$, $z=1$ and $z=2$.
The top panels show $\langle\Gamma_{\rm flare}\rangle$
(flares per unit time, per all galaxies, in the galaxy rest-frame).
The black, solid curves show all binaries embedded in Toomre-stable discs.
The blue, dashed curves show the flaring rate from only the binaries
whose circumbinary discs are luminous in the V-band,
and the red, dotted curves show those that are ultra-dim in the V-band
in between flares. This was determined by evaluating whether the redshifted
peak frequency in the spectral emission of the circumbinary
disc was higher or lower than $10^{14}~\Hz$ (equation \ref{eq:nupeak2a}).
The Figure shows that if all SMBH pairs merge via migration in a disc
with a central cavity, then the event rate could exceed $10^{-3}~\yr^{-1}$,
compared to the TDE rate estimates of $\Gamma_{\rm TDE}\sim 10^{-5}~\yr^{-1}$ \citep{MagorrianTremaine99}.
In other words, if $\sim 1\%$ of SMBH pairs produced cavity-accretion flares,
then their global event rate could be comparable to the rate of stellar TDEs.
(However, as mentioned in \S\ref{subsubsec:disc2},
recently merged systems may preferentially by obscured by dust.)
The overall fraction of flares occurring in systems with
optically dim circumbinary discs
increases with $z$ as the peak disc frequency is redshifted.

The bottom panels of Figure \ref{fig:pop} show the upper limits
for the fraction of all galaxies that host an accreting SMBH binary
embedded in a Toomre-stable circumbinary disc.
I have also plotted the maximum fraction of galaxies
hosting SMBH binaries with periods
$1~{\rm mo}<P_{\rm obs}<1~{\rm yr}$ (dotted red curves),
$1~{\rm yr}<P_{\rm obs}<10~{\rm yr}$ (short-dashed blue curves), and
$10~{\rm yr}<P_{\rm obs}<100~{\rm yr}$ (long-dashed purple curves).
The figures suggest that flaring binaries should typically have orbital periods
(flare frequencies) of several months to several years if the binary mass
is less than $\sim 10^8~\Msol$, and $\ga 10~{\rm yr}$ for $M\sim 10^9~\Msol$
(coincident, incidentally, with the flare period of OJ 287).

These upper limits are, of course, highly speculative, given the
large uncertainties regarding the formation rate and evolution of SMBH binaries,
the fraction of these objects that form prograde circumbinary accretion discs,
and the fraction of \textit{those} objects that are unobscured.
These calculations do, however, demonstrate the possibility that
cavity-accretion flares from SMBH binaries might be a common occurrence,
perhaps comparable to the incidence of luminous AGN activity.
The existence of a populationof flaring binaries could be hidden by the fact
that they may resemble inactive galaxies in between flares.
They could be discovered, alongside TDEs and other extragalactic
transients, in upcoming surveys such as \textit{LSST} and \textit{eROSITA}.

\section{Discussion}
\label{sec:discussion}

Above, I have argued that an accreting SMBH binary with
a prograde circumbinary disc may produce periodic X-ray/UV/optical
flares at cadences of years or decades, that this may occur
in nuclei that are optically quiescent in between outbursts,
and that such events may be common in the transient sky.
That is good and well as a piece of educated speculation,
but the inverse problem---that is, supposing that such flares are observed,
then discerning whether the cause is an accreting SMBH binary---is a much more difficult one.

It is well-known that most AGN exhibit aperiodic optical/UV fluctuations
at the $\sim 10\%$ level over timescales of weeks to years
\citep{Ulrich+97, Giveon+99, CollierPeterson01}.
However, examples in which the AGN appears to ``turn on''
briefly from a previously quiescent state are rare (e.g. \citealt{Gilli+00}, \citealt{GTB01}).
Transient AGN-like activity with rapidly rising and decaying optical and UV light curves
are generally thought to be stellar TDEs \citep{KomossaGreiner99, KomossaBade99, Gezari+06, Esquej+08, Gezari+09, Cappelluti+09, vanVelzen+11, Cenko+12, Saxton+12}.

Indeed, the flares proposed in this work are physically similar to TDEs, 
in that an amount of gas comparable to that of a star
is deposited close to a SMBH (which may previously appear quiescent)---except
that whereas the ``fuel'' in TDEs is deposited at a distance of a few
gravitational radii, the gas streams leaked from the circumbinary
disc are typically deposited at thousands of gravitational radii
(i.e., $\ltsim 0.1 a$).
One could also think of the flares as a  gas-rich analogue
of G2, the purported $\sim 10^{-5}~\Msol$ gas cloud\footnote{
\cite{Murray-ClayLoeb12} and \cite{ScovilleBurkert13}
suggest that G2 is instead a low-mass star with an associated disc or outflow.
}  falling toward   Sgr  A* \citep{Gillessen+12}.

The task of disentangling TDEs from SMBH binaries
is further complicated by the fact that the two phenomena are not mutually exclusive.
On the contrary, it has been suggested that SMBH binaries---or
a recently merged binary \citep{KomMer08,StoneLoeb11, StoneLoeb12}---may even enhance
the rate of TDEs, without the aid of
an accretion disc \citep{Chen+09, Chen+11, WeggBode11}.
Conversely, an accretion disc around a single SMBH
may enhance the TDE rate without a central binary \citep{KarasSubr07}.
Thus, the observation of multiple TDE-like transients
in a single galactic nucleus will not be enough to claim
that the cause is an accreting SMBH binary,
even if corroborated by additional circumstantial
indicators such as morphological evidence of a recent galaxy merger.

Fortunately, the cavity-accretion scenario proposed here has several basic predicted features
(other than periodicity) that should help to distinguish flaring supermassive binaries from TDEs.
The most significant prediction is that the flares can be produced by SMBHs
with $M_\bullet>{\rm few}\times 10^8~\Msol$, which are thought not to disrupt solar-type stars.
Another is that it requires the presence of a circumbinary accretion disc
with a cavity, which should have an anomalously red or optically dim
SED that may be observed in the infrared.
The cavity may also result in unusually weak broad emission lines in between flares.
AGN with such properties are known to constitute a small fraction ($\sim 1\%$)
of the general population \citep{Gibson+08, Shemmer+09}.

I have only discussed the thermal emission from the
optically thick accretion flow. However, it is widely established that SMBH accretion
can exhibit a wide range of emission features.
The diversity of emission is shared by TDE candidates---
e.g., the contribution of the outflowing wind \citep{SQ09,SQ11},
the production of relativistic jets \citep{Burrows+11, Bloom+11, Zauderer+11}
and associated radio transience \citep{GianniosMetzger11}.
\cite{Sesana+12} discussed the possibility that SMBH binaries may produce
periodic variability in the Fe K$\alpha$ emission line
(but this may be difficult to observe if the accretion flows inside the cavity
are short-lived);
the rapid onset of a hypercritical accretion flow may trigger
radiatively driven jets or outbursts \citep{TM10}.
The cavity-accretion flares discussed in this paper, if they occur in nature,
should have a similarly diverse variety of observable signatures.

In constructing the toy model of the accretion flare,
I made several order-of-magnitude parameterizations and estimates,
based on previous studies where possible,
for quantities such as the mass deposited by each stream
and the distance from the SMBH where the gas circularizes and begins to accrete.
The actual light curves and SEDs will also depend on the detailed thermal
and viscous evolution of the gas, including the possible onset of instabilities.
The emission signatures presented here are based on idealized
treatments of the accretion geometry and viscosity, and should
be viewed only as a proof-of-concept demonstration
based on order-of-magnitude normalizations.
Finally, the toy model assumes that the fueling accretion flow
forms on a short timescale much shorter than the binary orbital timescale;
cavity-accretion flares may develop more gradually
if the fueling flows are slow to form and circularize,
and this could help to distinguish them from TDEs.
The validity of the above theoretical treatments can be tested by
future numerical simulations.

\section{Conclusion}
\label{sec:concl}

In this paper, I explored the possibility that the periodic
leakage of circumbinary gas onto a SMBH binary produces
luminous transient flares that may interrupt intervals of apparent quiescence.
The underlying mechanism behind this hypothesis
consists of two characteristics of prograde accretion discs found
in numerous simulations: (i) the presence of a central, circumbinary
cavity in a prograde accretion disc, which reduces the emergent flux of the system
at frequencies above the optical;
and (ii) the periodic, punctuated leakage of the disc gas into the cavity,
which would trigger UV and soft X-ray emission as it accretes onto
one or both SMBHs.
The main findings for these ``cavity-accretion flares'' from SMBH binaries
are as follows:
\begin{enumerate}
\item The hypothesis that accreting SMBH binaries
produce quasiperiodically recurring, transient luminous flares
is based on the comparison of the depletion timescale of the
gas that enters the cavity with the binary's orbital period.
As long as the streams are circularized inside a radius comparable
to the SMBH's Hill sphere, the orbital energy available to
heat the gas is sufficiently large so that the
the post-shock accretion flow should have a 
viscous accretion timescale shorter than the binary's orbital period.
The post-shock accretion flow is expected to be
marginally geometrically thin and strongly advective,
similar to the slim disc class of accretion-disc solutions.
As is generally the case with such optically thick accretion,
the thermal emission of the flare
is expected to peak in the UV or soft X-rays, and extend to optical frequencies.
\item One or two transient flares may be produced per binary orbit, depending on whether one
or both SMBHs develop optically thick accretion flows.
\item Unlike TDEs, these flares can be powered by SMBHs of arbitrary mass,
including those in the $10^9~\Msol$ class.
\item Because the presence of the cavity reduces the high-frequency
thermal emission of the circumbinary disc, in between the outbu rsts
such a system may appear as an anomalously coloured AGN
or be mistaken for a quiescent galactic nucleus.
The circumbinary disc may also have unusually weak broad-line or X-ray emission.
\item The flares and the intravening quiescence
could manifest themselves as periodic optical outbursts,
such as those exhibited by OJ 287.
The cavity-accretion interpretation of OJ 287
differs from the disc-impact model proposed by
\cite{Valtonen+06} in several ways:
the required SMBH mass is smaller $(~10^9~\Msol)$;
 the optical dimness in between outbursts
is a direct consequence of the cavity, as opposed to reddening by dust;
 whereas the disc-impact model strongly constrains
the flare timing based on the binary orbit solution,
temporal stochasticity is expected in the cavity-flare model due
to the accretion dynamics.
One or both optical flares may be associated with a radio-loud
episode, depending on the spins of the SMBHs.
As stated above in (iii), the initial shock that circularizes the accretion flow
in the cavity-flare model could explain
the ``precursor flare'' events observed in this system.
\item If the gas sound speed in the post-shock accretion flow
scales with its Keplerian velocity (i.e., if its thermal profile is virial),
then the corresponding $\alpha$ viscosity prescription
will drive a central accretion rate that decays as $\dot{M}\propto t^{-4/3}$.
As with TDEs ($\dot{M}\propto t^{-5/3}$ canonically),
the monochromatic and bolometric light curves are not
expected to follow a simple power-law.
Thus, it may not be possible to deduce the mass accretion rate
from the light curve.
\item Most flaring SMBH binaries will have periods of $\sim 1-100~\yr$,
if the production of flares is contingent upon the Toomre stability
of the circumbinary disc.
More massive systems are capable of more luminous outbursts having
longer cadences. For binaries with $M\ltsim 10^6 ~\Msol$, the orbital periods
may be months or shorter; however, such low-mass binaries
may not produce flares because the cavity may close \cite{Kocsis+12II}.
\item Depending on the fraction of galaxy mergers that
produce sub-pc SMBH binaries with prograde circumbinary discs,
the global rate of cavity-accretion flares may be high enough to
be detected serendipotously by existing and future deep,
high-cadence observatories such as \textit{eROSITA} or \textit{LSST}.
Under the most liberal assumptions---that most galaxy mergers
result in the formation of such binaries and that the flares
are not preferentially obscured due to the recent merger of the hosts---the
global rate of cavity accretion flares from SMBH binaries may exceed the theoretical rate of TDEs.
\end{enumerate}

If they occur in nature, cavity-accretion flares from SMBH binaries
should be discovered by future surveys such as \textit{LSST} and \textit{eROSITA}
alongide TDEs and other extragalactic transients.
In the meantime, additional constraints may be placed on the prevalence of this mechanism
by monitoring (candidate) TDE host galaxies for recurrent activity.
The results presented here reinforce the notion
that accreting SMBH binaries may be a common source of
luminous astrophysical transients \citep{HKM09, THM10}.

\section*{ACKNOWLEDGMENTS}

I am grateful to the organizers of the
\textit{Tidal Disruption Events and AGN Outbursts} workshop,
held at the European Space Astronomy Centre in Madrid, Spain,
where the idea for this paper originated;
some preliminary results that led to this paper are presented
in the proceedings \citep{Tanaka12}.
I thank Linda Strubbe and Nicholas Stone for encouraging discussions;
Stefanie Komossa for bringing to my attention the radio
flaring properties of OJ 287;
Alister Graham for pointing out the slope of the
SMBH-host correlations at high masses; Zolt\'an Haiman and Kristen Menou for comments on the manuscript;
and the anonymous referee for suggestions that improved the clarity of this paper.

\bibliographystyle{mn2e}

\end{document}